\documentclass[]{aa520}
\usepackage{graphicx}
\usepackage{tabularx}

\usepackage{txfonts}
\begin{document}
    \title{A quadruply imaged quasar with an optical Einstein ring
    candidate: 1RXS~J113155.4$-$123155 \thanks{Based on data collected
    at the European Southern Observatory, La Silla, Chile}}

    \subtitle{}

    \author{D. Sluse\inst{1,2}, J. Surdej\inst{1}\thanks{Directeur de
    recherches honorifique du F.N.R.S. (Belgique)},
    J.-F. Claeskens\inst{1}, D. Hutsem\'ekers
    \inst{1,2}\thanks{Chercheur qualifi\'e du F.N.R.S. (Belgique)},
    C. Jean \inst{1}, F. Courbin \inst{1}, T. Nakos \inst{1,2,3},
    M. Billeres \inst{2}, \and S.V. Khmil \inst{4}}

    \offprints{sluse@astro.ulg.ac.be}

    \institute{Institut d'Astrophysique et de G\'eophysique,
    ULg, All\'ee du 6 Ao\^ut 17, B5C, B-4000 Sart
    Tilman (Li\`ege), Belgium \and European Southern Observatory,
    Alonso de Cordova 3107, Santiago 19, Chile \and Royal Observatory
    of Belgium, Avenue Circulaire 3, B-1180 Bruxelles, Belgium \and
    Astronomical Observatory of Shevchenko University, 3 Observatorna
    st., Kyiv UA-04053, Ukraine}

    \date{Received: ;  accepted: }

\abstract{We report the discovery of a new quadruply imaged quasar
surrounded by an optical Einstein ring candidate. Spectra of the
different components of 1RXS~J113155.4$-$123155 reveal a source at $z=$
0.658. Up to now, this object is the closest known gravitationally
lensed quasar. The lensing galaxy is clearly detected. Its redshift is
measured to be $z=$ 0.295. Additionally, the total V magnitude of the
system has varied by 0.3~mag between two epochs separated by 33
weeks. The measured relative astrometry of the lensed images is best
fitted with an SIS model plus shear. This modeling suggests very high
magnification of the source (up to 50 for the total magnification) and
predicts flux ratios between the lensed images significantly different
from what is actually observed. This suggests that the lensed images
may be affected by a combination of micro or milli-lensing and dust
extinction effects.

 }

    \titlerunning{Discovery of a quadruply imaged quasar at z=0.658}
    \authorrunning{Sluse et al.}

   \maketitle

%

\section{Introduction}

The peculiar and complex morphology of the source 1RXS~J113155.4$-$123155
(hereafter J1131GL) has been serendipitously unveiled during
polarimetric imaging of a sample of radio quasars carried out in May
2002 at ESO, La Silla. These observations are reported in
Section~\ref{sec:Diap} together with additional optical imaging
obtained in December 2002. Astrometry and photometry of the
gravitational lens system are also described. In
Section~\ref{sec:spec}, we present spectroscopic observations of the
source, the lens and a nearby companion (hereafter
J1131b). Section~\ref{sec:mod} is devoted to a simple lens model and
Section~\ref{sec:end} summarizes why this new gravitational lens is a
particularly interesting one. We have adopted throughout the paper
$H_0=$ 65~km s$^{-1}$ Mpc$^{-1}$, $\Omega_0=$ 0.3 and $\lambda_0=$
0.7.


\section{Direct imaging, astrometry and photometry} 
\label{sec:Diap}

Direct imaging of J1131GL has been obtained at two different epochs
with respectively EFOSC-2 at the 3.6~m telescope and EMMI-Red at the
3.5~m New Technology Telescope (NTT) at the La Silla observatory. On
May 2, 2002, we resolved the four components of J1131GL on a set of
2$\times$4 polarimetric images (corresponding to 4 different
orientations of the Half Wave Plate and 2$\times$150~s integration time
per orientation) taken through a combined V-band + Wollaston
prism. The average seeing measured on the frames is 1.1\arcsec\ and
the pixel size is 0.158\arcsec. Additional V and R images of J1131GL
have also been obtained under poor seeing conditions (1.6\arcsec) on
December 18, 2002. The coadded exposure time amounts to 480~s in V
and 960~s in R. The pixel size is 0.166\arcsec.

\begin{figure*} [tb]
\centering
{\includegraphics[bb=36 398 577 570, width=15.4cm]{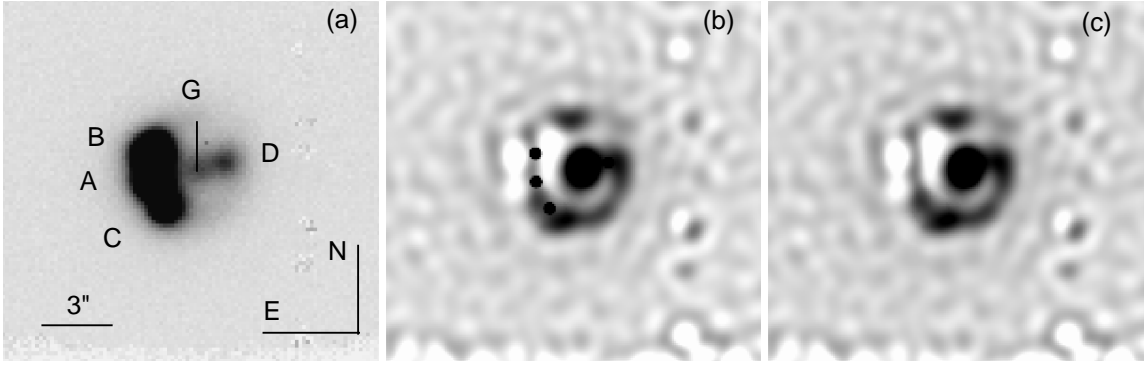}}
\caption{{\bf(a)} V direct image of J1131GL obtained with EFOSC-2 
(4$\times$150~s integration time). The 4 lensed images and the
deflecting galaxy are identified on this CCD frame. {\bf(b)}
Deconvolved image using the MCS method (see text) and {\bf(c)} Idem
but with the quasar images removed. A conspicuous Einstein ring
candidate is seen on the deconvolved images.}
\label{fig:lens1}
\end{figure*}

\begin{table*}[htb]
\centering
\caption{ {\it Left}: Relative positions between the different lensed 
components (B, C, D) and the lensing galaxy (G) with respect to
component A as deducted from the best seeing EFOSC-2 frame using the
GENERAL (GEN) and MCS codes (see text). {\it Right}: Relative
magnitude of the B, C, D lensed images with respect to A deducted with
the MCS method at two different epochs. The last two columns give
respectively the amplification $\mu_{\textrm{\tiny{exp}}}$ (sign = parity) predicted
by the SIS$+\gamma$ model and the corresponding relative magnitude
with respect to A (Section~\ref{sec:mod}).}
\begin{scriptsize}
 \label{tab:res}
\begin{tabular}{r|cccc|ccccc}
\hline \hline
ID & \multicolumn{2}{c}{$\Delta \alpha \cos \delta('') (J2000)$} & \multicolumn{2}{c|}{$\Delta \delta ('') (J2000)$ } 
&  V (02/05/2002) & V (18/12/2002) & R (18/12/2002) & $\mu_{\textrm{\tiny{exp}}}$ & $\Delta m_{\textrm{\tiny{exp}}}$ \\
\hline
& GEN & MCS & GEN & MCS & \multicolumn{3}{c}{MCS} & \multicolumn{2}{c}{model} \\
\hline
A &\multicolumn{2}{c}{$11 h 31 m 51.6 s$} &\multicolumn{2}{c|}{$-12\degr 31\arcmin 57\arcsec$} & $0$ & $0$ & $0$ &  $-25.78$ &$0$ \\
B & $+0.024\pm0.002$ & $+0.029\pm0.003$ &$+1.201\pm0.002$ &$+1.189\pm0.004  $  & $-0.45\pm0.04$ & $-0.46\pm0.06$ & $-0.49\pm0.06$ & $14.67$ & $+0.61$\\
C & $-0.563\pm0.003$ & $-0.573\pm0.002$ &$-1.062\pm0.003$ &$-1.124\pm0.003$ &$+0.62\pm0.07$  & $+0.62 \pm 0.08$  & $+0.57\pm0.08$ &$14.68$ & $+0.61 $\\
D & $-3.122\pm0.010$ & $-3.088\pm0.036$ &$+0.884\pm0.010$ &$+0.821\pm0.013$ &$+2.14\pm0.11$ & $+2.16 \pm 0.14$ & $+1.83\pm0.14$ &$-1.53$ & $+3.07$\\
G& $-1.898\pm0.015$ & $-1.911\pm0.034$ &$+0.559\pm0.015  $ & $+0.542\pm0.023$& - & -& - & - & - \\
\hline 
\end{tabular}

\end{scriptsize}
\end{table*}

\subsection {Image analysis}
\label {subsec:analysis}

By means of the GENERAL image decomposition program developed by Remy
et al. (\cite{remy}), we have fitted the EFOSC images of J1131GL (see
Fig.~{\ref{fig:lens1}a}) using 4 free adjustable PSF components,
altogether with a galactic light profile for the lens. Due to
significant distortions of the PSF across the field,
we have finally obtained the best results using a purely analytical 2D
Moffat PSF. 

Fitting a (non seeing-convolved) de Vaucouleurs or exponential disk
profile allows to roughly remove the low frequency signal of the
lensing galaxy and to unveil a ring-like structure at a level higher
than 3$\sigma$ above the noise. This feature is spectacularly
confirmed by the deconvolution of the images using the MCS code
(Magain et al. \cite{magain}): a non uniform ring passing through the
4 point-like components is seen on the deconvolved images
(Figs.~{\ref{fig:lens1}}b,c).
\noindent Additionally, we measured that the integrated signal from the 
gravitational lens system (A-D and G) was not significantly polarized
(i.e. total polarization $\sim$ 0.2$\pm$0.08\%).
\noindent GENERAL and MCS have also been applied to the EMMI frames 
but despite of a more stable PSF across the field, the poorer seeing
prevented us to reach better results than with EFOSC-2.

\subsection {Relative astrometry and photometry}

Astrometry of the lensed components (B-D) and of the lensing galaxy
(G) relatively to A has been derived using the GENERAL and MCS
codes. With the MCS algorithm, the position of the lens G has been
measured by deconvolving a point-like source at the lens position.
The results listed in Table~{\ref{tab:res}} were derived for the best
seeing EFOSC frame. They are in statistical agreement with the
positions retrieved for the second epoch.
\noindent The image deconvolved by the MCS code is a sum of analytical point
sources {\it and} of a diffuse numerical deconvolved background. Due
to this, the ring and the lensing galaxy are better taken into account
with the MCS code rather than with GENERAL. Consequently, the
resulting photometry of the point-like components obtained with MCS is
more reliable. The results obtained for the two epochs are reported in
Table~{\ref{tab:res}}.
\noindent The 1$\sigma$ errors listed in Table~{\ref{tab:res}} are formal 
errors on the fit for the results derived with GENERAL. In the case of
the MCS code, they reflect the dispersion of the results when changing
the initial conditions and deconvolution parameters (i.e. smoothing,
position, flux and background step in the $\chi^2$ fit). Systematic
errors on the photometry and on the astrometry are probably not
negligible for D and G due to their small separation and their equally
low S/N.

\subsection {Absolute photometry and variability}

The two polarimetric standards HD~155197 and HD~298383 have been
observed with the same setting (i.e. V band + Wollaston) during the
photometric night on May 2, 2002. These stars allowed us to calculate
a zeropoint of 25.85$\pm$0.03~mag.  Consequently, the integrated V
magnitude of the system is estimated to be 16.63 and $V_{\textrm{\tiny{A}}}=$
17.97$\pm$0.09 (deducted with MCS). The brightest parts of the ring
have a surface brightness $V\sim$ 23.9~mag arcsec$^{-2}$. The
estimated mean surface brightness of the galaxy inside a 4\arcsec\
radius is $V\sim$ 22.7~mag arcsec$^{-2}$ and its integrated magnitude
inside the same radius is V=18.4.
\noindent Since the conditions were not photometric during the December
observations, we performed {\it differential} photometry of the
integrated system J1131GL with respect to various objects in the
field. Due to the polarimetric nature of the EFOSC frames (i.e. the
field is splitted in non contiguous bands), only 4 objects could be
used for this purpose; and it was necessary to also use the 10~s
acquisition frame. We found that the integrated flux of the system
(A-D and G) was brighter by 0.29$\pm$0.04~mag in December 2002. Since
the relative photometry between the 4 components is quite similar at
both epochs (see Table~{\ref{tab:res}}), intrinsic variability is
very likely responsible for this difference.

\section{Low resolution spectroscopy with EMMI}
\label {sec:spec}

Two sets of low resolution spectra (2\arcsec\ slit) have been obtained
using the new CCD on the EMMI-Red arm at the NTT: {\bf (1)} at the end
of May 2002, we obtained a 900~s unresolved spectrum (PA $\sim$
15$^{\circ}$, from North to East) of the three bright components A, B,
C with the CCD in the 1$\times$1 bin mode and {\bf (2)} on January 25,
2003, we obtained 2$\times$900~s spectra with the CCD in the
2$\times$2 bin mode (corresponding to 3.58~$\AA$ pixel$^{-1}$) and the
slit passing through A, D, G and J1131b, located at 25\arcsec\
East of J1131GL. Standard bias subtraction, flatfielding and
spectrum extraction procedures were used. Wavelength calibrations of
the spectra were done with an He-Ar lamp. Due to the absence of a
spectrophotometric standard in January 2003, we used the
spectrophotometric standards LTT~1788 and LTT~2415 observed with a
5\arcsec\ slit in December 2002 to correct these data.

\begin{figure}[tb]
\centering
\resizebox{\linewidth}{!}{\includegraphics[bb=51 51 407 300]{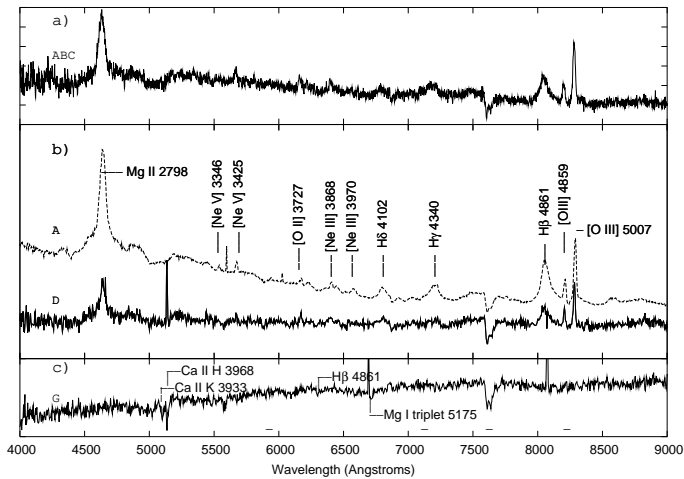}}
\caption{{\bf (a)} Integrated spectrum of the three
bright components A, B, C. {\bf (b)} Spectrum of A (2$\times$900~s) and
D (900~s). Emission lines of a quasar at $z=$ 0.658 are easily
identified.  {\bf (c)} Spectrum of the lensing galaxy (900~s). The
typical absorption lines of an elliptical galaxy at $z=$ 0.295 are
identified. The flux scales are arbitrary. The small horizontal lines
above the wavelength axis on (c) indicate telluric absorption lines.}
\label{fig:spectra}
\end{figure}

The integrated spectrum of A, B, C obtained in May 2002
(Fig.~{\ref{fig:spectra}}a) shows a continuum and emission lines
typical of a redshifted quasar. We could separate these three
components using the deconvolution algorithm based on a Maximum
Entropy Method developed by Khmil and Surdej ({\cite{khmil}}). The
separated spectra of A, B, C obtained by this algorithm were found to
be very similar with slight possible changes in the continuum slope
but they also remained highly correlated due to the poor seeing of
1.7\arcsec. The second set of spectra obtained in January 2003 under
an average seeing of 0.66\arcsec\ have completed our knowledge of this
system. Due to the spatial sampling of 0.332\arcsec pixel$^{-1}$, the
spectra of A, G and D slightly overlapped. Nevertheless a careful
choice of the apertures enabled us to extract the spectrum of A on
both frames and the spectra of G and D on the second one. An average
spectrum of A is shown in Fig.~{\ref{fig:spectra}}b. We identify on
this spectrum typical emission lines of a quasar at a redshift
z=0.658$\pm$0.001. This redshift was determined by fitting a Gaussian
on the MgII $\lambda$ 2798~\AA\ and [OIII] $\lambda\lambda$ 4959,
5007~\AA\ emission lines. Unfortunately, the 2\arcsec\ slit slightly
scatters the light from B and C (located less than 1.2\arcsec\ away
from A) and consequently the spectrum of A is contaminated on each
exposure by these two components.

Because of the overlapping and of the similar brightness of G and D
their spectra are mutually contaminated. We thus subtracted one
spectrum from the other after adequate scaling in order to visually
minimize the contamination (i.e. to remove quasar emission lines from
G and the 4000~\AA\ break from A). This handicraft process suggests a
relative reciprocal contamination smaller than 10\%. The resulting
spectra are shown in Figs.~{\ref{fig:spectra}}b-c. Component D shows
similar emission lines as A, but has a flatter slope, not generated by
decontamination. The spectrum of G is typical of an elliptical galaxy
(Kennicutt \cite{kenni}) showing absorption lines redshifted at $z=$
0.295$\pm$0.002 (e.g. CaII K\&H $\lambda\lambda$ 3933, 3968~\AA, G
band $\lambda$ 4304~\AA, etc).

The spectrum of J1131b shows Balmer absorption lines and a continuum
typical of an A type star contaminated by a background galaxy lying at
2\arcsec\ from this object. This star was previously identified by
Bauer et al. (\cite{bauer}) as the most likely source of the X-ray
(ROSAT) and radio (NVSS) emission present in this field. They also
reported a redshift z=0.654 for this object. We firmly reject their
identification, mistakenly matched with the quasar we have observed
here.

\section{A simple lens model}
\label{sec:mod}

The relative positions of the quasar images and of the lensing galaxy
with respect to image A (see Table~\ref{tab:res}) have been fitted
using two simple lens models: the Singular Isothermal Ellipsoid (SIE,
Kormann et al. {\cite{kormann}}, Kassiola \& Kovner {\cite{kassiola}})
and the Singular Isothermal Sphere plus an external shear
(SIS$+\gamma$). The best fit is obtained with the SIS$+\gamma$ model
and yields the following results: the angular Einstein radius
$\theta_E =$ 1.819$\pm$0.006\arcsec, the shear $\gamma=$
0.123$\pm$0.003 and the shear position angle $\varphi=$
14.84$\pm$0.11$^\circ$ (from North to East). The uncertainties on the
parameters come from the fit of the model on 1000 Monte-Carlo
synthetic observations compatible with the observed errors. The
direction orthogonal to the shear axis $\varphi$ does not point
towards any bright object close to the lens. The reduced ${\chi}^{2}$
(for 3 degrees of freedom) is significantly smaller for the
SIS$+\gamma$ (${\chi}^{2}_{\gamma}=$ 19) than for the SIE lens model
(${\chi}^{2}_{\textrm{\tiny{SIE}}}=$ 203) because the latter cannot
reproduce correctly the observed lens position. If one does not fit
the lens galaxy position, the reduced $\chi^2$ is comparable for the
SIE and the SIS$+\gamma$ but the SIE model predicts a lens position
0.4\arcsec\ away from the observed one. Finally, using the formulae of
Witt et al. ({\cite{witt}}) for singular isothermal lens models with
shear, we predict the time delays between B and the other multiple
images: $\tau_{\textrm{\tiny{BC}}}=$ 0.01~d,
$\tau_{\textrm{\tiny{BA}}}=$ 0.87~d and $\tau_{\textrm{\tiny{BD}}}=$
96~d. The leading image is B and the time delay sequence is thus BCAD.

    \begin{figure}[tb]
   \centering
\resizebox{\linewidth}{!}{\includegraphics{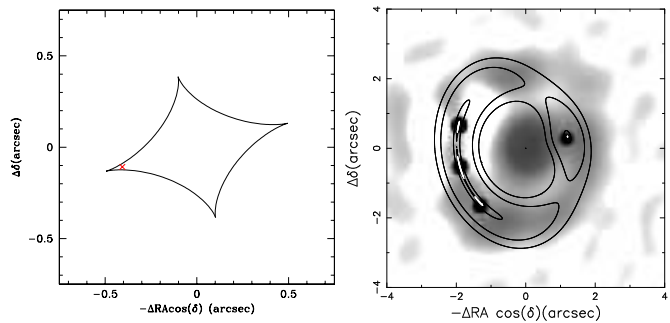}}
\caption{Results of the SIS$+\gamma$
model expressed in a system of coordinates centered on the lensing
galaxy G. {\it Left panel:} the source plane position of the source
(x), relative to G, is ($-$0.411\arcsec,$-$0.1084\arcsec). This
corresponds to a closest distance to a fold caustic of 0.016\arcsec\
(115 pc in the source plane). {\it Right panel:} the contours
(calculated from the model) corresponding respectively to 110, 740,
3700 and 5200 pc from the central engine in the source plane are
overplotted on the deconvolved EFOSC frame. }
\label{fig:model}
    \end{figure}

As can be seen from Fig.~{\ref{fig:model}}, this system is a long axis
quad with a source located very close to a cusp (at about
0.1\arcsec). This allows very high magnification of the host galaxy at
a few hundreds of pc from the center. The latter is the likely origin
of the optical ring-like structure joining the lensed point-like
images. The two minima images of the arrival time surface, B and C
(with positive parity), should have about half the flux of the central
one (the saddle point A), as demonstrated by Mao (\cite{mao}) and by
Schneider and Weiss (\cite{schneider}). The observed B/A magnification
ratio (see Table~\ref{tab:res}) does not follow this generic
prediction: B is even the brightest image. Since the C/A ratio is
``normal'', and since the time delays are very short between A, B, C,
we suspect B to be affected by micro/milli-lensing (D could also
be). Even if the V-band and R-band flux ratios are quite similar (see
Table~\ref{tab:res}), we cannot exclude that dust extinction also
plays a role in the observed flux ratios. The unlensed absolute
magnitude of the source is $M_{\textrm{\tiny{B}}}=$ $-$22.7 if we
conservatively take ${\mu}_{\textrm{\tiny{A}}}=$ 10 and $B-V=$
0.2. Thus, the source is, strictly speaking, an AGN/Seyfert~1.

\section{Discussion and conclusions}
\label{sec:end}

Direct imaging and long slit spectroscopy of 1RXS~J113155.4$-$123155
presented in this letter have enabled us to show that this object is a
quadruply imaged quasar ($V_{\textrm{\tiny{ind}}}
\in [17.5,20.1]$; $\Delta\theta\sim$ 1.2\arcsec) at redshift
$z=$ 0.658$\pm$0.001 lensed by an elliptical galaxy at $z=$
0.295$\pm$0.002. The MCS method has made possible to enhance the
signature of an Einstein ring candidate barely seen on the direct
images. The simple modeling by an SIS $+\gamma$ shows that the source
is located very close to a cusp allowing the host galaxy to cross
the caustic and to generate the ring seen on the deconvolved
images. As it is observed in many quads, there is a discrepancy
between the observed image flux ratios (especially between the saddle
point A and the minimum B images) and the ones predicted by
modeling. This may reflect the necessity to use a lens model involving
a small percentage of substructures (Schechter and Wambsganss
{\cite{schechter}}, references therein). More data on J1131GL are
necessary before drawing any definite interpretation of the observed
discrepancy. Nevertheless, the explanation of the flux ratios in this
system should probably involve micro/milli-lensing and/or dust
extinction. Only individual spectra obtained simultaneously from UV to
NIR (and taken at time intervals equal to the time-delays) for each
component will enable one to disentangle between these effects. 
Finally, we have shown that the integrated flux has varied by 0.3 mag
between May and December 2002.

The source J1131b located at 25\arcsec\ East of J1131GL was wrongly
identified by Bauer et al. (\cite{bauer}) as the most likely optical
counterpart of the X-ray (Voges et al. \cite{voges}) and radio (Condon
et al. \cite{condon}) emission present in this region. Our
identification of this object as a hot star suggests that J1131GL
(already cataloged with the {\it RXS} notation) is the true source of
the X-ray and radio emission. Note that the discovery of a
gravitationally lensed system in a multi-wavelength survey is not
surprising due to the large expected multi-band magnification bias
(Borgeest et al. {\cite{borgeest}}, Wyithe et al. {\cite{wyithe}}).

This first set of data suggests the necessity of good spatial
resolution and high signal to noise ratio multi-wavelength imaging
(from radio to X-ray) in order to use the rare characteristics of
this system as many observational constraints for an accurate
modeling. In this framework, the shape of the Einstein ring is an
invaluable asset to determine independently the shape of the lens
potential and of the unlensed source (Kochanek et al. {\cite
{kochanek}}). Moreover, a precise lens inversion of the lens
equation (e.g. Warren and Dye
\cite{warren}) should enable one to draw a unique multi-wavelength
picture of the source and to retrieve information on the source at
angular scales inaccessible with present day and even future
instrumentation.

This bright system brings {\it together} rare properties
(i.e. quad, bright optical Einstein ring, small redshift, high
amplification), nearly unique among the known gravitational
lens systems. These features make 1RXS~J113155.4$-$123155 a very promising
astrophysical laboratory for future investigations, including the
possibility for an independent determination of the Hubble parameter
$H_{0}$ based on time delay measurements.


\begin{acknowledgements}
Our research was supported in part by PRODEX (Gravitational lens
studies with HST), by contract IUAP P5/36 ``P\^ole d'Attraction
Interuniversitaire" (OSTC, Belgium) and by the ``Fonds National de la
Recherche Scientifique" (Belgium). F.C. is supported by the European
commission through Marie Curie Fellowship MCFI-2001-0242. The
collaborative grant EROS/CONICYT C00405 between Chile and France is
also acknowledged. The referee, D.Rusin, is warmly acknowledged for
his constructive remarks on the first draft of this letter. We finally
want to thank Y. Naz\'e for her help with the use of the WIP software.
\end{acknowledgements}

\end{document}